\documentclass[preprintnumbers,amsmath,amssymb,floatfix,superscriptaddress,nofootinbib,12pt]{revtex4}
\usepackage{graphicx}
\usepackage{epsfig}
\usepackage{bm}
\usepackage{amsfonts}
\def\be {\begin{equation}}
\def\ee {\end{equation}}
\def\ba {\begin{eqnarray}}
\def\ea {\end{eqnarray}}
\begin{document}

\title{Phantom Wormholes in (2+1)-dimensions}
\author{Mubasher Jamil}
\email{mjamil@camp.nust.edu.pk} \affiliation{Center for Advanced
Mathematics and Physics,\\ National University of Sciences and
Technology, Rawalpindi, 46000, Pakistan}
\author{M. Umar Farooq}
\email{m_ufarooq@yahoo.com} \affiliation{Center for Advanced
Mathematics and Physics,\\ National University of Sciences and
Technology, Rawalpindi, 46000, Pakistan}

\begin{abstract}
\textbf{Abstract:} In this paper, we have constructed a
(2+1)-dimensional wormhole using  inhomogeneous and anisotropic
distribution of phantom energy. We have determined the exact form of
the equation of state of phantom energy that supports the wormhole
structure. Interestingly, this equation of state is linear but
variable one and is dependent only on the radial parameter of the
model.
\end{abstract}
\maketitle

\newpage

A typical wormhole is characterized by a tunnel in spacetime
connecting  two arbitrary spacetime sections. These sections could
either belong to the same spacetime or to two different spacetimes.
The wormhole geometry arises naturally as a solution of the Einstein
field equations \cite{visser,jamil2,jamil1}. Interest in wormhole
physics was initiated when Morris and Thorne investigated the
wormhole structure and proposed that the material required to
construct it has to be exotic, i.e. its (negative) radial pressure
and energy density must satisfy the inequality $|p|>\rho$
\cite{morris}. They also concluded that this structure could also
serve as a time travel machine if it is horizon-free.

From the cosmological perspective, a candidate for exotic matter
exists namely the phantom energy. Presently it is well-motivated
from the observational data that the observable universe is pervaded
with the phantom energy, which is characterized by
$\omega=p/\rho<-1$ \cite{caldwell,caldwell1}. In recent years,
several studies are performed regarding construction of wormholes
with the use of phantom energy as an exotic matter
\cite{andrew,sushkov,zas,peter,cataldo}. The phantom energy is
exotic due to its weird and esoteric properties: its energy density
increases as the universe expands; its accretion onto all
gravitationally bound objects results in disassociating them; it can
rip apart the spacetime itself in a finite time which is called the
Big Rip.

Understanding the nature of gravity is one of the hardest and most
challenging problems in theoretical physics. Interest in
(2+1)-dimensional theories of gravity - especially general
relativity - dates back to early sixties. Since then several toy
models have been built up in (2+1) gravity which help in
understanding the corresponding (3+1) dimensional problems
\cite{carlip}. Gravity in (2+1) dimensions behaves very differently
compared to the usual (3+1) dimensional gravity, for example, the
gravity does not exist outside the matter source and remains
confined locally. Since gravity does not propagate outside the
gravitating source, gravitational waves don't arise in this case. In
recent years, models of wormholes in (2+1) dimensional gravity are
presented \cite{farook,rubic,kim,mann}. In these models, coordinate
$\theta$ is fixed so that $d\theta=0$. Consequently, this reduces
the complexity of the field equations.

The metric of a (2+1)-dimensional Morris-Thorne (MT) wormhole is
given by \cite{farook,rubic,kim} \be
ds^2=-e^{2f(r)}dt^2+\frac{1}{1-\frac{b(r)}{r}}dr^2+r^2d\phi^2, \ee
where $f(r)$ is called potential function while $b(r)$ is the shape
function. These functions are arbitrary functions of radial
coordinate $r$ and will be determined below for a specific choice of
matter distribution. The radial coordinate has a range that
increases from a minimum value at $r_0$, corresponding to the
wormhole throat, and extends to infinity. For the wormhole to be
traversable, conditions commonly termed stability and
traversability, are imposed on these two functions, namely: $f(r)$
must be bounded for all values of $r$; $b^\prime(r_0)<1$ at $r=r_0$;
$b(r)<r$ for all $r>r_0$ and $b/r\rightarrow0$ as
$|r|\rightarrow\infty$. The stress energy components in an
orthonormal frame of reference are $T_{00}=\rho$, $T_{11}=p$ and
$T_{22}=p_{t}$. Here $\rho$ is the energy density, $p$ is the radial
pressure while $p_{t}$ is the transverse pressure.

The Einstein field equations become (units are $c=1=G$)
\begin{eqnarray}
\rho (r)&=&\frac{b^{\prime }r-b}{16\pi r^{3}},\\
p(r)&=&\frac{(1-b/r)f^{\prime }}{8\pi r},\\
p_{t}(r)&=&\frac{(1-b/r)}{8\pi }\Big[f^{\prime \prime }-\frac{(b^
{\prime }r-b)}{2r(r-b)}f^\prime +f^{\prime 2}\Big].
\end{eqnarray}
Here prime ($\prime$) denotes differentiation with respect to $r$.
The energy conservation equation is obtained by evaluating
$T^{AB}_{;A}=0$, with $A,B=0,1,2$. It gives
\begin{equation}
p^{\prime }+f^{\prime }\rho +\Big(f^{\prime }+\frac{1}{r}\Big)p-\frac{p_{t}}{r}=0.
\end{equation}
Note that in the above equations (3) to (5), the function $f$ must
not be a constant (i.e. $f'\neq0$) otherwise the field equations
become identically zero. Below we shall determine explicit form of
$f$ for a specific choice of two parameters. To solve the field
equations, we choose the following ansatz for the shape function and
pressures
\begin{equation}
b(r)=\frac{b_{0}}{r^m},\ \ m=0,1,2,...
\end{equation}%
\begin{equation}
p_{t}=\alpha p,
\end{equation}%
where $\alpha $ and $b_{0}$ are constants. Notice that $\alpha$ is
dimensionless while $b_0$ possesses dimensions of $(length)^{m+1}$.
It is easy to check that Eq. (6) satisfies the stability conditions
for the wormhole. The second ansatz (7) says that the ratio of
transverse to radial pressure will remain constant although both can
vary differently. Our task is to find $\rho$, $f$, $p$ and $p_{t}$
using Eqs. (2) to (7).

Inserting $(6)$ in $(2)$, we have
\begin{equation}
\rho(r) =-\frac{b_{0}(m+1)}{16\pi r^{m+3}}.
\end{equation}%
Since $\rho $ is always positive so we require $b_{0}<0$. Making
use of Eqs. (2) to (7) and after simplification, we arrive at
\begin{equation}
f^{\prime \prime
}+\Big[\frac{b_{0}(m+1)}{2r(r^{m+1}-b_{0})}-\frac{\alpha
}{r}\Big]f^{\prime }+f^{\prime 2}=0.
\end{equation}
To solve this equation, we rewrite it as \be
\frac{f^{\prime\prime}}{f^\prime}-\frac{m+1}{2}\Big(\frac{1}{r}-\frac{r^m}{r^{m+1}-b_0}\Big)
-\frac{\alpha}{r}=-f^\prime.  \ee Integrating it we get \be f^\prime
e^f=\frac{Cr^{\alpha+\frac{1}{2}+\frac{m}{2}}}{\sqrt{r^{m+1}-b_0}},
\ee where $C$ is a constant of integration and for the sake of
convenience we fix $C=1$. Integration once more leads to
\begin{equation}
f(r)=\ln \left[\frac{2r^{\frac{1}{2}(3+m+2\alpha)}
\sqrt{1-\frac{r^{1+m}}{b_0}}{}_2F_{1}\Big(\frac{ 3+m+2\alpha
}{2+2m},\frac{1}{2},\frac{5+3m+2\alpha
}{2+2m},\frac{r^{1+m}}{b_0}\Big)}{\sqrt{ r^{1+m}-b_0}(3+m+2\alpha
)}\right].
\end{equation}%
Here ${}_2F_{1}$ is a hypergeometric function representing a series
expression. It should be noted that logarithmic form for $f(r)$ have
been obtained for a MT wormhole in (3+1)-dimensions as well
\cite{peter}. Also notice that $f(r)$ does not give a finite value
as $r\rightarrow\infty$, so the solution is not asymptotically flat.
Hence we may match this interior solution to an exterior vacuum
spacetime at a junction radius $R$ \cite{tanvi}. Notice that in
(2+1) dimensions, the only exterior vacuum solution is the
stationary BTZ spacetime \cite{btz,btz1}, given by
\begin{equation}
ds^2=-\Big(-M+\frac{r^2}{l^2}\Big)dt^2+\Big(-M+\frac{r^2}{l^2}\Big)^{-1}dr^2+r^2d\phi^2.
\end{equation}
Here $M$ corresponds to mass of the spherically symmetrical object
while $l=1/\sqrt{-\Lambda}>0$ and $\Lambda<0$ is the cosmological constant.
In order to match the interior wormhole solution with the exterior BTZ solution,
 we impose the continuity of the metric coefficients, $g_{AB}$, across a surface $S$, i.e.
\begin{equation}
g_{AB}^{\text{int}}|_S=g_{AB}^{\text{ext}}|_S.
\end{equation}
The wormhole metric is continuous from the throat radius $r=r_0$ to a finite distance $r=R$.
 Explicitly Eq. (14) can be written as
\begin{eqnarray}
g_{00}^{\text{int}}|_S=g_{00}^{\text{ext}}|_S,\\
g_{11}^{\text{int}}|_S=g_{11}^{\text{ext}}|_S.
\end{eqnarray}
Notice that $g_{22}$ is already continuous, so we don't need any matching equation
 for it. The last two equations yield respectively
\begin{eqnarray}
\frac{2R^{\frac{1}{2}(3+m+2\alpha)}
\sqrt{1-\frac{R^{1+m}}{b_0}}{}_2F_{1}\Big(\frac{ 3+m+2\alpha
}{2+2m},\frac{1}{2},\frac{5+3m+2\alpha
}{2+2m},\frac{R^{1+m}}{b_0}\Big)}{\sqrt{ R^{1+m}-b_0}(3+m+2\alpha
)}&=&-M+\frac{R^2}{l^2},\\
1-\frac{b_0}{R^{1+m}}&=&-M+\frac{R^2}{l^2}.
\end{eqnarray}
Here $M$ now refers to the mass of wormhole. Using (6) and (12) in (3), we get
\begin{equation}
p=-\frac{b_0r^{-(3+m)}\sqrt{1-\frac{r^{1+m}}{b_0}
}(3+m+2\alpha)}{16\pi{}_2F_{1}\Big(\frac{ 3+m+2\alpha
}{2+2m},\frac{1}{2},\frac{5+3m+2\alpha
}{2+2m},\frac{r^{1+m}}{b_0}\Big) }.
\end{equation}
Putting (19) in (7), we obtain
\begin{equation}
p_{t}=-\frac{b_0r^{-(3+m)}\sqrt{1-\frac{r^{1+m}}{b_0}
}\alpha(3+m+2\alpha)}{16\pi{}_2F_{1}\Big(\frac{ 3+m+2\alpha
}{2+2m},\frac{1}{2},\frac{5+3m+2\alpha
}{2+2m},\frac{r^{1+m}}{b_0}\Big) }.
\end{equation}%
Alternatively, Eq. (20) can be obtained by inserting (6) and (12) in (4).
In Fig. 1 and 2, we have plotted the magnitudes of the radial and the transverse
 pressures against the radial coordinate. These show that both the pressures have
  arbitrary large values near the throat while these vanish in the asymptotic limit
   of $r$. This shows that the matter distribution also satisfies the condition of
asymptotic flatness, consistently with the wormhole geometry.    The
difference of radial and transverse pressures represents surface
    tension which plays very crucial role in compact stars \cite{jamil}. Fig. 3
     shows that the behavior of surface tension is analogous to the two pressures.

Comparison of Eqs. (8) and (19) yields a relationship between
pressure and energy density, given by
\begin{equation}
p=\left[\frac{\sqrt{1-\frac{r^{1+m}}{b_0}
}(3+m+2\alpha)}{(m+1){}_2F_{1}\Big(\frac{ 3+m+2\alpha
}{2+2m},\frac{1}{2},\frac{5+3m+2\alpha
}{2+2m},\frac{r^{1+m}}{b_0}\Big) }\right]\rho.
\end{equation}
On comparing Eq. (21) with $p=\omega\rho$, we get
\begin{equation}
\omega(r)=\frac{\sqrt{1-\frac{r^{1+m}}{b_0}
}(3+m+2\alpha)}{(m+1){}_2F_{1}\Big(\frac{ 3+m+2\alpha
}{2+2m},\frac{1}{2},\frac{5+3m+2\alpha
}{2+2m},\frac{r^{1+m}}{b_0}\Big) }.
\end{equation}
It shows that the wormhole under consideration satisfies a variable equation
 of state. The variable EoS arises naturally while solving the field equations
  for the wormhole. Interestingly a variable EoS unifies various forms of dark
   energy including phantom energy and Chaplygin gas, both of which support
   the wormhole spacetime \cite{peter1}. The behavior of $\omega(r)$ is given
    in Fig. 4 and it shows that the EoS parameter $\omega$ has to be negative
     to model a wormhole. It naturally yields negative radial pressure and
     positive energy density.

The case of isotropic pressure $p=p_{t}$ is obtained by fixing $\alpha=1$. We have
\begin{equation}
p=p_{t}=\left[\frac{\sqrt{1-\frac{r^{1+m}}{b_0}
}(5+m)}{(m+1){}_2F_{1}\Big(\frac{ 5+m }{2+2m},\frac{1}{2},\frac{7+3m
}{2+2m},\frac{r^{1+m}}{b_0}\Big) }\right]\rho.
\end{equation}
Similarly, the EoS parameter $\omega$ becomes
\begin{equation}
\omega(r)=\frac{\sqrt{1-\frac{r^{1+m}}{b_0}
}(5+m)}{(m+1){}_2F_{1}\Big(\frac{ 5+m }{2+2m},\frac{1}{2},\frac{7+3m
}{2+2m},\frac{r^{1+m}}{b_0}\Big) }.
\end{equation}
We would also comment that the case of vanishing pressure $p=0$ (dust) is not
 allowed in the present formalism since it will make $f(r)$ unbounded.
 The dust cases in the framework of braneworld wormholes are investigated
  in \cite{subenoy,lobo}.

The specific dimensionless parameter $\xi$, defined by $\xi=(p-\rho)/|\rho|$,
 characterizes how the exotic or normal matters are distributed around the
  wormhole's throat \cite{morris,kim}. The \textit{exoticity} at or near
  the throat of the wormhole is required to be non-negative, $\xi>0$. The
   positivity of the exoticity ensures that wormhole will satisfy the flare-out
    condition as well. From Eq. (15), the exoticity becomes
\begin{equation}
|\rho|\xi=-\frac{b_0(m+1)}{16\pi
r^{m+3}}\left[\frac{\sqrt{1-\frac{r^{1+m}}{b_0}
}(3+m+2\alpha)}{(m+1){}_2F_{1}\Big(\frac{ 3+m+2\alpha
}{2+2m},\frac{1}{2},\frac{5+3m+2\alpha
}{2+2m},\frac{r^{1+m}}{b_0}\Big) }  -1\right].
\end{equation}

In Fig. 5, we have plotted the exoticity against the radial coordinate. It shows
 that the exoticity remains positive while it converges to zero for large $r$.
  Hence the wormhole is surrounded by the exotic phantom energy, right from its
   throat to a sufficiently large distance.
This also suggests that we can construct a wormhole with a sufficiently large
radius that could be traversable for two dimensional beings.

In summary, our objective in this article has been
to present a mathematical prescription for obtaining a
wormhole in low dimensional spacetime. The wormhole is supported by
an external source of phantom energy which is anistropically distributed.
 The wormhole spacetime satisfies a variable equation of state, which is
 in good agreement with earlier models available in the
 literature. It is also shown that the otherwise asymptotically
  non-flat wormhole could be converted to an asymptotically flat
  one by matching the various components of its metric with the exterior BTZ spacetime.

\newpage
\begin{figure}
\includegraphics{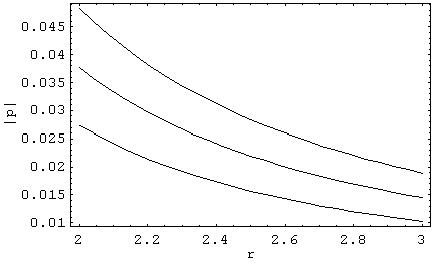}\\
\caption{The radial pressure is plotted against radial coordinate
while other parameters are fixed at $b_0=-1$, $m=2$ and
$\alpha=1,2,3$ (top to bottom).}
\end{figure}
\begin{figure}
\includegraphics{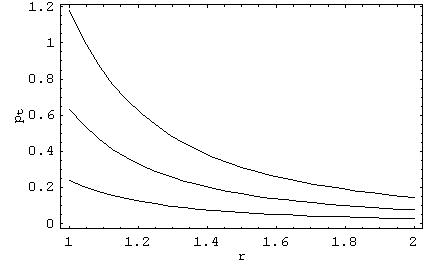}\\
\caption{The transverse pressure is plotted against radial
coordinate while other parameters are fixed at $b_0=-1$, $m=2$ and
$\alpha=1,2,3$ (top to bottom).}
\end{figure}
\begin{figure}
\includegraphics{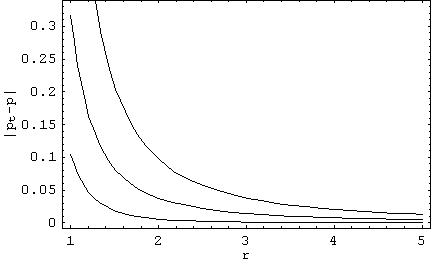}\\
\caption{The surface tension is plotted against radial coordinate while
 other parameters are fixed at $b_0=-1$, $m=2$ and $\alpha=-3,2,3$ (top to bottom).}
\end{figure}
\begin{figure}
\includegraphics{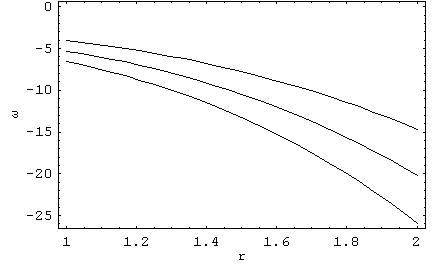}\\
\caption{The equation of state parameter $\omega$ is plotted against radial
 coordinate while other parameters are fixed at $b_0=-1$, $m=2$ and $\alpha=1,2,3$
  (top to bottom).}
\end{figure}
\begin{figure}
\includegraphics{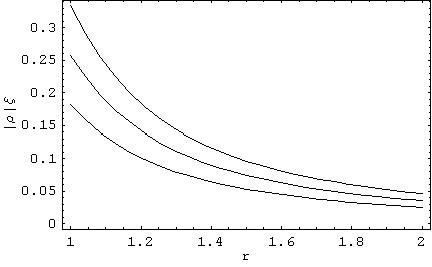}\\
\caption{The exoticity $|\rho|\xi$ is plotted against radial
coordinate while other parameters are fixed at $b_0=-1$, $m=2$ and
$\alpha=1,2,3$ (top to bottom).}
\end{figure}

\end{document}